\documentclass[a4paper,12pt]{article}
\usepackage{epsfig}
\begin{document}

\begin{center}
{\Large \bf Quark Probability Distribution at Finite Temperature and Density }\\[5mm]
{Lianyi He, Guang Bian, Jinfeng Liao and Pengfei Zhuang}\\[4mm]
{\small{\it Physics Department, Tsinghua University, Beijing 100084, China}\\[5mm]}
\today\\[5mm]
\end{center}

\begin{abstract}
\setlength{\baselineskip}{16pt} Quark deconfinement phase
transition at finite temperature and density is investigated in
the frame of quantum mechanics. By solving the Schr\"odinger
equation for a heavy quark in a thermal mean field, we calculate
the quark probability distribution as a function of temperature
and density. The confined wave function in the vacuum expands
outward rapidly when the temperature and density are high enough.
The obtained phase transition line agrees qualitatively with the
result of lattice QCD.
\end{abstract}
\vspace{0.3in}

\setlength{\baselineskip}{18pt}

It is widely accepted that the collective effect of a multiparton
system can change the vacuum structure of Quantum Chromodynamics
(QCD)\cite{ftft}, and the partons confined in a hadron bag can
move out when the temperature and density are high enough and form
a new state of matter, the so called quark-gluon plasma
(QGP)\cite{qgp}. This kind of new matter may be produced in the
early stage of a high-energy heavy-ion collision\cite{rhic}. The
dense partons, produced in the vacuum at high temperature or
compressed at high baryon density, make the quark wave functions
overlap in a space with dimension of the colliding nuclei. An
important question in relativistic heavy ion collisions is how to
identify the new state of matter if it is created in the early
stage. According to the numerical calculation of lattice
QCD\cite{lattice}, the critical temperature of quark deconfinement
is about 150 MeV when the density effect is excluded.

The idea that the overlap of wave functions, or the space
extension of wave functions at finite temperature and density is
the original reason of quark deconfinement phase transition arises
naturally an interesting question to study the behavior of quark
distribution function in the frame of quantum mechanics. In this
letter, we will study a quark moving in a thermal mean field
describing multiparton interactions, and obtain the quark
probability distribution by solving the Schr\"odinger equation at
finite temperature and density. The temperature and density at
which the quark wave function expands outward rapidly can be
considered as the critical point of quark deconfinement. Because
of the limitation of non-relativistic quantum mechanics, we focus
on the motion of a heavy quark at finite temperature and density,
corresponding to, for example, the dissociation of
$J/\psi$\cite{jpsi} as a signature of QGP in relativistic heavy
ion collisions.

In the string-like models\cite{rhic} that describe the quark
confinement well, one uses a linear potential $V=kr$ to express
the interaction between two quarks. To simplify the numerical
calculation in the following, we use a three dimensional square
well
\begin{eqnarray}
\label{v0}
V(r) = \left\{\begin{array}{ll}
0 & 0<r<a\\
V_0 & a<r<b\\
\infty & r>b \end{array}\right.
\end{eqnarray}
instead of the linear potential. Here $a$ is the boundary of the
region where the motion of the quark is asymptotically free, and
can be taken as the radius of the hadron constructed by the
considered quarks, $b$ is the space scale of the system, and $V_0$
should be so large that the quark wave function can be confined in
the asymptotically free region $r,a$. Since the region of
collective interaction in high-energy heavy-ion collisions is of
the order of the colliding nuclei, we treat $b$ as the radius of a
heavy nucleus. In our numerical calculations, we take the
parameters of the confinement potential as $a=1$ fm, $b=5$ fm, and
$V_0 = 10$ GeV.

At finite temperature and density, the quark will be affected by
many-body interactions in the region $r>a$ in addition to the
confinement potential. In mean field approximation, the
thermodynamic potential of the quasi-particle system is\cite{ftft}
\begin{equation}
\label{omega}
\Omega(T,\mu) = -g\int{d^3{\bf p}\over
(2\pi)^3}{p^2\over 3E_q}\left(f_q(T,\mu)+f_{\bar q}(T,\mu)\right)
\end{equation}
with the  quark and antiquark densities
\begin{eqnarray}
\label{density}
f_q(T,\mu) &=& {1\over e^{E_q-\mu\over T}+1}\ ,\nonumber\\
f_{\bar q}(T,\mu) &=& {1\over e^{E_q+\mu\over T}+1}\ ,
\end{eqnarray}
where $g$ is the quark degree of freedoms of flavors, colors and
spins, $T$ and $\mu$ are respectively the temperature and baryon
chemical potential, $E_q = \sqrt{m_{eff}^2+p^2}$ is the quark
energy with effective mass $m_{eff}$ obtained in the mean field.
Because the contribution from heavy quarks to the thermodynamic
potential is very small compared with light quarks, we will only
take $u$ and $d$ quarks into account, i.e., $g=12$.

From the rules of thermodynamics, the pressure of the system is
\begin{equation}
\label{pressure}
P(T,\mu) = -\Omega(T,\mu)\ .
\end{equation}
Considering the relation between the potential $V_{th}$ and the
force $\vec F_{th}$ that the heavy quark we study feels in the
region $r>a$,
\begin{equation}
\label{force} \vec F_{th}(\vec r) = -\nabla V_{th}(\vec r)\ ,
\end{equation}
i.e.,
\begin{equation}
\label{force2}
-{dV_{th}(r)\over dr} = -4\pi r^2 P(T,\mu)\ ,
\end{equation}
and assuming that the potential falls down with increasing $r$ and
vanishes on the boundary of the system, namely $V_{th}(r=b)=0$,
the potential $V_{th}$ can be expressed as
\begin{equation}
\label{vth} V_{th}(r|T,\mu) = -\int_r^b dr' 4\pi r'^2P(T,\mu) =
{4\over 3}\pi(b^3-r^3)\Omega(T,\mu)\ .
\end{equation}
With the obtained potential, the motion of the heavy quark at
finite temperature and density is described by the Schr\"odinger
equation
\begin{equation}
\label{schro} i{\partial\over\partial t}\psi(\vec r,t) =
\left(-{1\over 2m}\nabla^2+V_{eff}(r|T,\mu)\right)\psi(\vec r,t)\
,
\end{equation}
where $m$ is the mass of the heavy quark, and $V_{eff}$ is the
effective potential defined as
\begin{eqnarray}
\label{veff} V_{eff}(r|T,\mu) = \left\{\begin{array}{ll}
0 & 0<r<a\\
V_0+V_{th}(r|T,\mu) & a<r<b\\
\infty & r>b \end{array}\right.
\end{eqnarray}
By separating the variables of the wave function $\psi(\vec r,t) =
R(r)Y(\theta,\varphi)e^{-iEt}$, and letting $\phi(r)=rR(r)$, we
obtain the radial equation for a stationary state
\begin{equation}
\label{phi} -{1\over 2m}{d^2\phi(r)\over
dr^2}+\left(V_{eff}(r|T,\mu)+{l(l+1)\over 2mr^2}\right)\phi(r) =
E\phi(r)\ ,
\end{equation}
where $E$ is the eigenvalue of the quark energy. Since we focus in
this letter on quark deconfinement at finite temperature and
density, we discuss in the following the ground state of the
Schr\"odinger equation only. The corresponding radial equation is
then simplified as
\begin{equation}
\label{phi0} -{1\over 2m}{d^2\phi_0(r)\over
dr^2}+V_{eff}(r|T,\mu)\phi_0(r) = E_0\phi_0(r)\ .
\end{equation}

We first discuss qualitatively the deconfinement phase transition
through the behavior of the effective potential as a function of
$T$ and $\mu$. From the $T$- and $\mu$-dependence of the
thermodynamic potential, we can immediately obtain the conclusion
that the confinement potential in the vacuum at $T=\mu =0$ will be
suppressed by the thermal mean field $V_{th} <0$, and the motion
of the quark will extend outside when the temperature and density
increase. Since the mean field decreases monotonously from the
maximum on the boundary $r=a$ to the minimum on the other boundary
$r=b$, the suppression at $r=a$ is most strong, and the potential
at $r=b$ does not change at finite temperature and density.

In order to illustrate quantitatively the above estimated quark
deconfinement phase transition, we should solve the Schr\"odinger
equation (\ref{phi0}) at finite temperature and density to get the
quark probability distribution
\begin{equation}
\label{pho} \rho(r|T,\mu) \propto r^2R^2(r|T,\mu)\ .
\end{equation}
We take the considered heavy quark in the Schr\"odinger equation
as a $c$ quark which forms $J/\psi$ with $m=1.6$ GeV, and the
effective mass of the light $u$ and $d$ quarks in $V_{eff}$ as
their constituent quark mass $m_{eff}=0.35$ GeV.

\begin{figure}[ht]
\vspace*{+0cm} \centerline{\epsfxsize=7cm  \epsffile{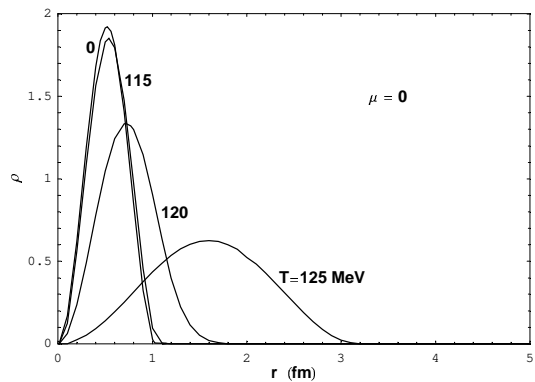}}
\caption{\it The quark probability distribution at finite
             temperature for $\mu=0$.} \label{fig1}
\end{figure}

\begin{figure}[ht]
\vspace*{+0cm} \centerline{\epsfxsize=7cm  \epsffile{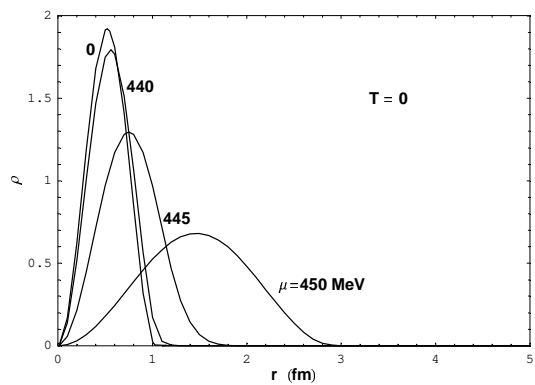}}
\caption{\it The quark probability distribution at finite
             chemical potential for $T=0$.} \label{fig2}
\end{figure}

In the limit of $\mu = 0$, the probability distribution as a
function of temperature obtained by solving numerically the
Schr\"odinger equation (\ref{phi0}) is shown in Fig.(\ref{fig1}).
In the vacuum with $T=\mu=0$, the chosen parameters for the square
well guarantee that almost the whole quark wave function is
restricted in the region $r<a$, that means quark confinement. As
temperature increases the quark motion extends outside very slowly
in the beginning, then the extension is accelerated rapidly in the
neighborhood of $T = 120$ MeV, which indicates that the two $c$
quarks forming $J/\psi$ begin to dissociate in the hot mean field,
and finally the quark moves in the whole region of the system
$0<r<b$. The case $T=0$ is similar to that of $\mu=0$, the quark
probability distribution as a function of $r$ for different
chemical potentials is indicated in Fig.(\ref{fig2}). The
probability expands outward rapidly in the neighborhood of $\mu =
445$ MeV.

\begin{figure}[ht]
\vspace*{+0cm} \centerline{\epsfxsize=6cm  \epsffile{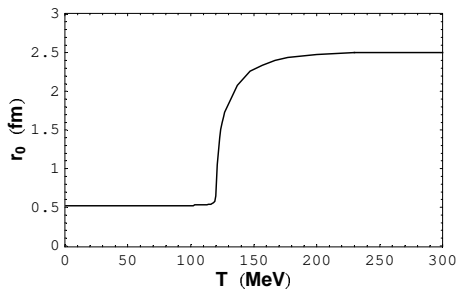}}
\caption{\it The most probable radius as a function of
             temperature for $\mu=0$.} \label{fig3}
\end{figure}

\begin{figure}[ht]
\vspace*{+0cm} \centerline{\epsfxsize=6cm  \epsffile{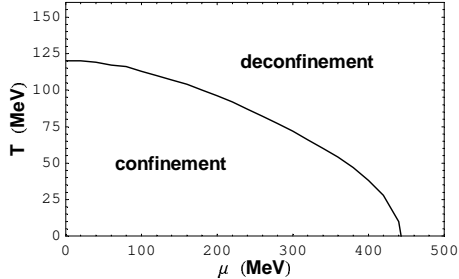}}
\caption{\it The phase transition line separating the region with
             quark confinement from that with quark deconfinement.}
\label{fig4}
\end{figure}

It is well known that for a particle moving in a central field the
most probable radius describes the bound degree of the stationary
state. The most probable radius $r_0(T,\mu)$ of the $c$ quark
motion at finite temperature and density defined by
\begin{equation}
\label{r0} {\partial \rho(r|T,\mu)\over \partial r}|_{r=r_0} = 0
\end{equation}
is indicated as a function of $T$ for $\mu = 0$ in
Fig.(\ref{fig3}). It is clear to see that around $T=120$ MeV $r_0$
goes up from $a/2$ to $b/2$ very fast. We can determine the
critical temperature $T_c$ and chemical potential $\mu_c$ of the
quark deconfinement phase transition as the point at which the
most probable radius $r_0(T_c,\mu_c)$ reaches $a$, the boundary of
the asymptotically free region in the vacuum,
\begin{equation}
\label{tcuc} r_0(T_c,\mu_c) = a\ .
\end{equation}
The critical line which separates the deconfinement phase from the
confinement phase is shown in Fig.(\ref{fig4}). While the most
probable radius $r_0$ jumps from $a/2$ to $b/2$ in a very narrow
region around $T_c$, it is still continuous at the critical point,
and the phase transition we study is a continuous one. When the
temperature and chemical potential exceed the critical values, the
region where the quark is asymptotically free will expand, i.e.,
the position of $a$ will move outward.

In summary, by describing the collective effect as a thermal mean
field and combining it with the confinement potential in the
vacuum, we have demonstrated the physical picture of quark
deconfinement in the frame of quantum mechanics through solving
the Schr\"odinger equation and calculating the quark probability
distribution at finite temperature and density. When the
temperature and density are high enough, the quark wave function
extends outward from inside the hadron. The critical temperature
and density around which the wave function extension happens
rapidly separates the confinement from deconfinement. The obtained
critical temperature $T_c = 120$ MeV at $\mu = 0$ agrees
qualitatively with the lattice QCD calculation.

{\bf Acknowledgments}: The work was supported by Grant Nos.
NSFC19925519, 10135030, and G2000077407.


\begin{thebibliography}{20}

\bibitem{ftft} J. I. Kapusta, {\it Finite Temperature Field Theory}, Cambridge University Press,
               1993; M. le Bellac, {\it Thermal Field Theory}, Cambridge University Press, 2000;
\bibitem{qgp}  {\it Quark-Glon Plasma}, ed. R.C.Hwa (World Scientific, 1990).
\bibitem{rhic} C. Y. Wong, {\it Introduction to High-Energy Heavy-Ion Collisions},
               World Scicetific, 1995.
\bibitem{lattice} F. Karsch, Nucl. Phys. {\bf A698}(2002)199c.
\bibitem{jpsi} T. Matsui and H. Satz, Phys. Lett. {\bf B178} (1986) 416;
               C. Gerschel and J. H\"ufner, Ann. Rev. Nucl. Part. Sci. {\bf 49} (1999) 255.
\end{thebibliography}
\end{document}